\documentclass[twocolumn,widetext,amsmath,amssymb]{revtex4}
\usepackage{graphicx}
\usepackage{dcolumn}
\usepackage{bm}
\usepackage{url}
\begin{document}

\title{Limited Urban Growth: London's Street Network Dynamics since the 18th Century}

\author{A. Paolo Masucci$^1$}
\author{  Kiril Stanilov$^{2,1}$}
\author{ Michael Batty$^1$}
\affiliation{1-Centre for Advanced Spatial Analysis, University College of London, W1T 4TJ, London, UK.\\
 2-Martin Centre, University of Cambridge, CB2 1PX, Cambridge, UK.}

\date{\today}

\begin{abstract}
We investigate the growth dynamics of  Greater London defined by the administrative boundary of the Greater London Authority,  based on the evolution of its street network during the last two centuries. 
 This is done by employing a unique dataset, consisting of the planar graph representation of nine time slices of  Greater London's road network spanning 224 years, from 1786 to 2010. 
 Within this  time-frame, we address the concept of the \textit{metropolitan area} or \textit{city} in physical terms, in that  urban evolution reveals observable transitions in the distribution of  relevant geometrical properties.
  Given that London has a hard boundary  enforced by its long standing \textit{green belt}, we show that its street network dynamics can be described as a fractal space-filling phenomena up to a capacitated limit, whence its growth can be predicted with a striking level of accuracy.
   This observation is confirmed by the analytical calculation of key topological properties of the planar graph, such as the topological growth of the network and its average connectivity.
   This study thus represents an example of a strong violation of  Gibrat's law. 
   In particular, we are able to show analytically how  London evolves from a more loop-like structure, typical of planned cities, toward a more tree-like structure, typical of self-organized cities. 
   These observations are relevant to the discourse on sustainable urban planning with respect to the control of urban sprawl in many large cities which have developed under the conditions of spatial constraints imposed by green belts and hard urban boundaries. 
  \end{abstract}
 \pacs{ 89.75.-k, 89.65.Lm, 89.75.Fb, 89.75.Da}

\maketitle


\section{Introduction}

Understanding spatio-temporal patterns in  complex  transportation systems is a major problem for efficient spatial organization.  \cite{westnat,banavar}. 
These systems span a wide range of natural and technological phenomena, from biology to urban systems. They are represented in studies such as leaf venation, crack pattern formation, river networks and urban street networks, for systems embedded in a two-dimensional Euclidean space, and ant galleries, circulatory systems, soap froths, and pipe networks for those embedded in a three-dimensional space  \cite{ants,bohn,bohn2007,klein}. 
Within transportation systems, a particularly relevant field is focused on studies of urban growth  \cite{maksenat,mikecc}. These are not simply paradigms of complexity, but they hold the key to many statistical regularities that have resided at the centre of scientific debate for many years, such as  Zipf's law for rank-size distributions and  Gibrat's law of proportionate growth  \cite{zipf,mikeclock,gibrat}.  

\textit{Planar graphs} are basic tools for understanding transportation systems embedded in two-dimensional space, in particular \textit{urban street networks}, where the street intersections are the vertices and the street segments connecting two intersections are the links  \cite{volk}. 
As these graphs are embedded in a two-dimensional surface, the \textit{planarity criteria} requires that the links do not cross each other. 
Planar graphs are the oldest graphs used in topological analysis  \cite{euler}, but their properties are still widely unknown due to difficulties arising from incorporating such planarity criteria into analytical calculation \cite{barth2011, masuccilond}. 
Although it is now well understood how the quest for transport optimality leads to the formation of reticulate networks rather than trees \cite{corson,magnasco}, disentangling the interconnections between topological and metrical properties for reticulate planar networks is still an open problem.

\begin{figure*}[ht]
\begin{center}
\centerline{\includegraphics[width=0.75\textwidth]{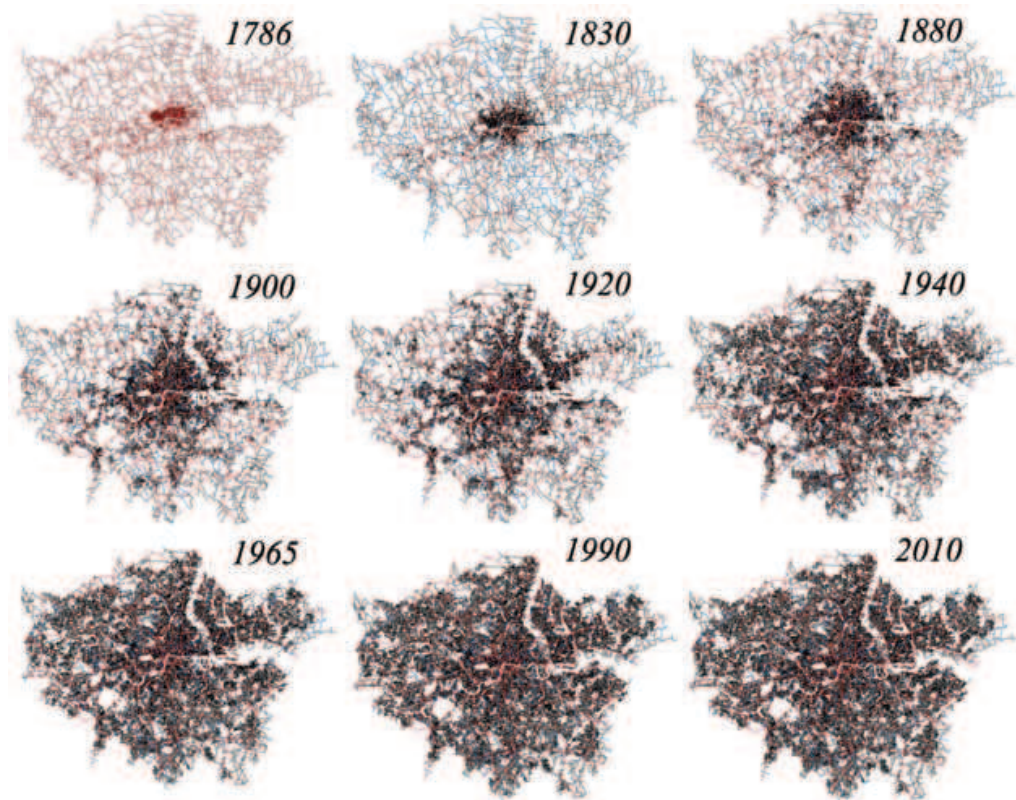}}
 \caption{  The  street network in the GLA (Greater London Area) from 1786 to 2010. Different   road colours correspond to different road classifications (red A roads and motorways, blue B roads, gray minor roads).}\label{f1}
\end{center}
 \end{figure*}
 
Here we analyse a unique dataset based on the street patterns of London defined as the Greater London Authority area (GLA hereafter) at nine time instants represented as nine map series spanning over 224 years - from 1786 to 2010 (see Fig.1). In these maps, each street segment is classified according to a four level hierarchy consisting of motorways, class A, class B, and minor roads, thus enabling us to extract the hierarchy of the network as recorded in the maps  without further assumptions. 
First, we speculate on the problem of the city's boundary. In this context, we show that the core of London's urban street network can be well defined by the statistical properties of the underlying street network, which are reflected by the  transitions in the distribution of certain geometrical properties.
Second, we show that the growth in the number of vertices and links of the planar graphs representing London can be treated as a fractal space-filling phenomena within a capacitated limit and thus described in terms of logistic functions through the Verhulst model \cite{logistic}. 
This observation allows  the London street network's growth problem to be treated analytically, and enables us to forecast with striking precision some key topological quantities about the street network dynamics. 

As a relevant outcome of this analysis, we highlight that the presented results represent a strong violation of the Gibrat's law, which states that urban growth is independent of city size \cite{gibrat}.
Moreover, these results allow us to forecast the evolution of the extent of the city and its sprawl \cite{irwin}, which has important implications for urban planning, and more generally provide a novel and efficacious approach to the study of transport systems embedded in two-dimensional space.

\section*{The Dataset}

A unique feature of the database employed by this project is the extensive time coverage, which spans over 200 years of London's urban growth. The data set that we developed includes the generation of time series maps for the 1,600 $km^2$ area defined by the present boundary of the Greater London Authority. Our time series captures in a chronological sequence the evolution of London's road network in 1786, 1830, 1880, 1900, 1920, 1940, 1965, 1990 and 2010. This allows us to trace the evolution of Greater London's road network from the incipient stages of metropolitan growth at the dawn of the industrial revolution to the present.

For  documentation of the network's evolution, we used a selection of highly detailed historic maps \cite{maps,maps1,maps2,maps3,maps4,maps5}, which allowed us to identify each of the existing roads at the time when the maps were created. We imported geo-referenced TIFF images of the historical maps in ArcGIS and traced manually the road centre lines on screen to create ArcGIS polylines. We excluded pedestrian paths and alleyways, but included the traditional London mews houses, which represent a substantial element of London's road network. 

On the road classification for the first two maps (1786 and 1830), the classification shown on the original maps has been used. These first two maps have highlighted what was called at that time ``principal roads".  An \textit{A road} classification has been assigned to all of the roads shown as principal roads on the 1786 map. The new principal roads which appear on the second map (1830)  have been given \textit{B road} classification.

For the remaining maps (1880-2010), which are all produced from Ordnance Survey maps, the current official classification of roads has been used. This method is  problematic to the extent that this classification has been introduced gradually since the 1920s and it is not  consistent across space and time - many of the roads have changed their classification over the years. Given this, we chose to use the present road classification (2010) and apply it backwards - in other words whenever a road appears (e.g. in 1940) it takes the class which it currently has (in 2010). However, we did not assign an A or B rank classification to loose fragments - meaning that for a road to be classified in these higher classes it has to be connected  at least to one end of a road of class A or B.

The dataset was then transformed into a weighted planar graph, where each intersection is a vertex $N=N(x,y)$ and each link is a street segment with a  weight given by its hierarchical classification.

Historical population data were extracted from \cite{demo}.

\begin{figure}[ht]
\begin{center}
\centerline{\includegraphics[width=0.5\textwidth]{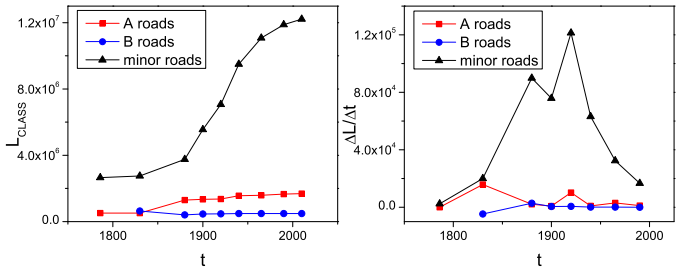}}
 \caption{  \label{f3c} Left panel: sum of the length of the street segments of a given class $L_{CLASS}(t)$ as a function of time expressed in meters. Right panel: growth rate $\frac{\Delta L}{\Delta t}$ for the  sum of the length of the street segments of a given class as a function of time expressed in meters/year. }
\end{center} 
 \end{figure}

\section*{The problem of city boundaries}\label{cb}

When analysing urban structure, we consider the city as being composed of many layers of infrastructure which underpin its social and economic functioning  \cite{stewart1958}.
 These are interconnected and coevolve, and lead to many different definitions of the city's physical extent. 
 Thus the definition of a \textit{city} can be quite blurred with respect to these layers. 
 Cities are usually analysed within their administrative boundaries, or within the extent of their urbanised area defined in terms of their population densities \cite{rozengib,rozenzipf}. 
 Nevertheless, a precise definition of a city's physical extent is crucial to any statistical analysis and extremely relevant when measuring fundamental relations, as for example in Gibrat's law and  Zipf's Law \cite{westsublin,rozengib,zipf}.
  Here we deal with a city, London, which has been capacitated by an artificial boundary  imposed to limit its growth and as such, it is representative of a number of  world cities such as Paris, S\~ao Paulo, Hong Kong, and Seoul, that are similarly constrained. 

 \begin{figure}[ht]
\begin{center}
\centerline{\includegraphics[width=0.5\textwidth]{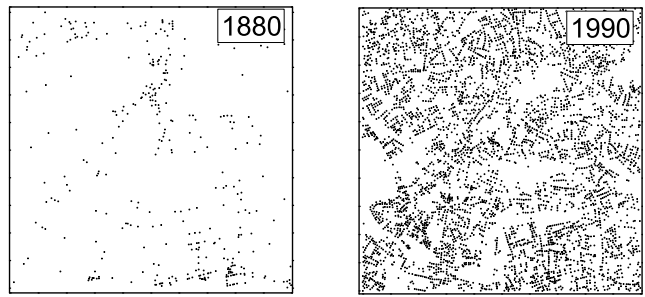}}
 \caption{\label{f3b} Left panel: Street intersections inside  a 10 km side square in the non-urbanized area of the GLA in 1880. Right panel:  Street intersections inside the same square  in 1990. }
\end{center}  
 \end{figure}   
 
City growth as a street network can be understood as the coevolution of two distinct phenomena, based on the hierarchy of its roads.
 On the one hand, we have the growth of major roads ( including  motorways, class A and class B roads) and, on the other, the growth of minor roads.
  A and B roads represent the backbone of the city, concentrating  the main flows of people and materials sustaining the city.
  Minor roads divide the blocks created by the A and B roads into smaller areas, and  are mainly devoted to local residential and business use \cite{pdual}. 
  
  The fact that the initial development of the A and B roads  generally precedes   the development of minor roads in the case of London's evolution  is quite clear from  Fig.1, where a relatively dense net of A and B roads is already present in 1786. The city then develops by filling the areas enclosed by A and B roads with minor roads.
    In Fig.\ref{f3c}, we show the measure of the sum of the lengths of the street segments of a given class $L_{CLASS}(t)$ as a function of time  in the left panel, and its change rate $\Delta L/\Delta t$ in the right for GLA's street network growth (motorways have been excluded from this analysis, since they only appear  from the 1965 map onwards and occupy a small number of transport links in Greater London).
 From this analysis, it is clear that the main structure of the A and B road  backbone pre-dated the city in 1786, although   additions to this backbone have increased by a factor of 4.2 during the subsequent 224 years. 
 However,  minor roads comprise most of the network growth that has taken place. 
This mixture  of major and minor roads leads to a considerable fragmentation of the major road system,  increasing the number of intersections. We believe that this mechanism ultimately   generates the kind of complex street pattern that we experience in large cities \cite{stevenet}.

 \begin{figure}[ht]
\begin{center}
\centerline{\includegraphics[width=0.5\textwidth]{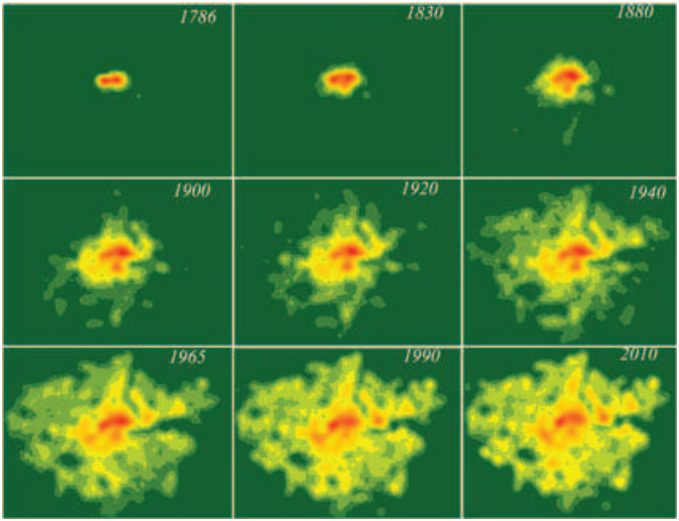}}
 \caption{\label{f2} Street intersection density surfaces in the GLA area from 1786 to 2010.}
\end{center} 
 \end{figure}

These two different phenomena are reflected in different street intersection patterns depicted in Fig.3, which shows a sample   10km x 10km  GLA area at a time interval of around one century. In the left panel, we show the intersection pattern of the non-urbanized area of the 1880 GLA map comprised of  $N=355$ points (intersections), which have an average density $\rho=3.55\cdot 10^{-6}m^{-2}$  . If the system were homogeneous, this density would suggest a length-scale of the order $\lambda\sim 1/\sqrt{\rho}\approx 530m$, one that we generally experience when we are outside the city's boundaries. In the right panel, we show the intersections for the same grid square in 1990, including  $N=4704$ intersections with an average density $\rho=4.7\cdot 10^{-5}m^{-2}$  .  This gives a scale to the system of the order of $\lambda\approx 146m$, that is a spatial scale   we experience typically in compactly developed cities. 

  \begin{figure}[ht]
\begin{center}
\centerline{\includegraphics[width=0.5\textwidth]{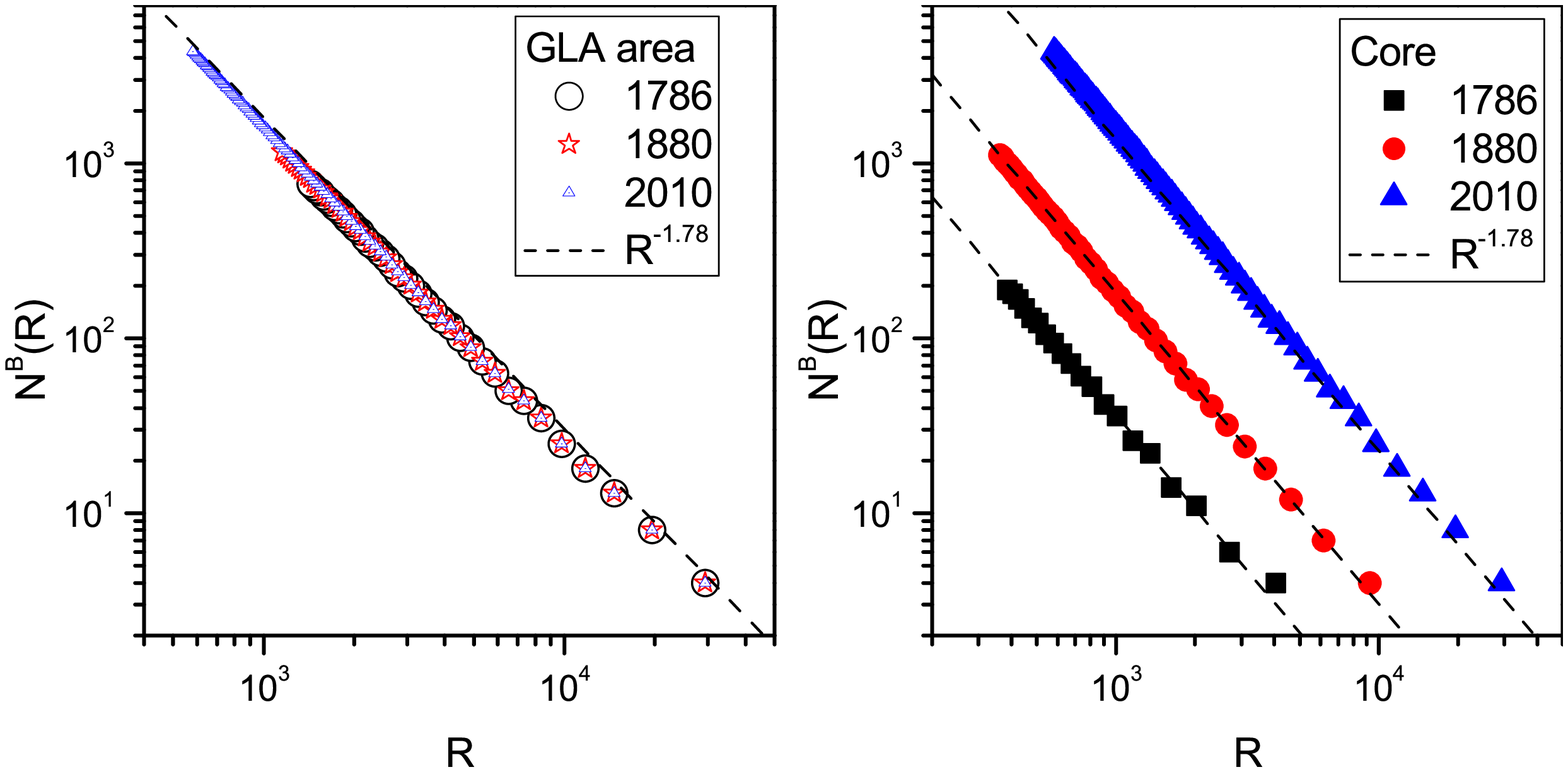}}
 \caption{\label{f3} Number of occupied squares $N^B(R)$ at scale $R$ for the intersection density maps and measurement of the fractal exponent $D_F$. Left panel: For the GLA. Right panel: For the density \textit{core} we define as \textit{London}. }
\end{center}  
 \end{figure}

These  observations relate the identification of an urban area to the street intersection density in that area. This idea was  introduced by  Jia et al. in \cite{Jiang}, bringing to the fore the concept of \textit{natural cities}.
In order to give strength to this reasoning, in Fig.\ref{f2} we show a qualitative analysis  of the intersection density in the GLA area. The figure is reminiscent of  a wild fire spreading, or more generally of a percolation phenomenon \cite{caldwildfires}. If we then use box-counting to calculate the number of boxes $N^B(R)$  at scale $R$  that are occupied by intersections, we know that this quantity scales as a power law  $N^B(R)\propto R^{-D_F}$ for fractal objects, where $D_F$  is the fractal dimension \cite{mikefract}.

 \begin{figure}[ht]
\begin{center}
\centerline{\includegraphics[width=0.5\textwidth]{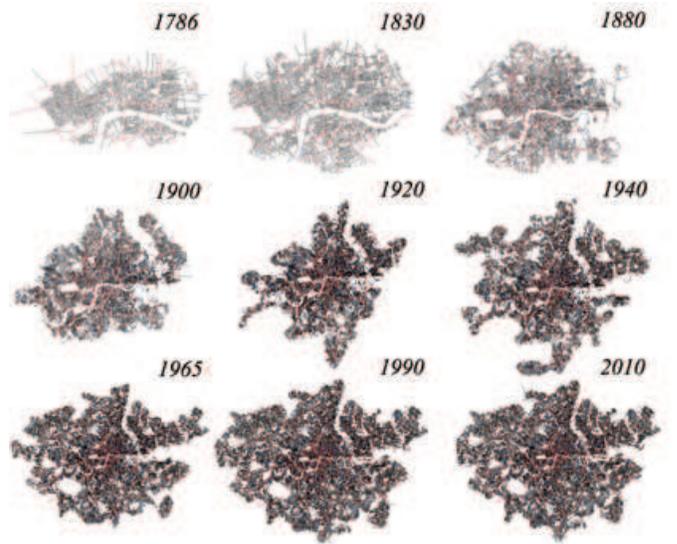}}
\caption{\label{f5} Our definition of London's street network,  as the urban core of the GLA  area derived  using the  Jenks clustering algorithm. }
\end{center}   
 \end{figure}

In the left panel of Fig.\ref{f3}, we show $N^B(R)$  for  street patterns in the GLA area at three time slices, each separated by about one century. The relationships overlap, closely fitting a power law with a fractal exponent  $D_F\approx 1.78$. The right panel of the same figure  displays the results of  an identical analysis but for the \textit{city} area of London (which we define  below as the \textit{core}). Interestingly, the behaviour of the road networks does not change at these different scales, which is a typical property of fractal objects, but the growth pattern is clearly different in the core from the GLA area. This analysis shows that the intersection density is a robust property of the road networks in the GLA. This observation allows us  to address the \textit{boundary of the city} or \textit{urban core} problem in the intersection density space.

  \begin{figure}[ht]
\begin{center}
\centerline{\includegraphics[width=0.5\textwidth]{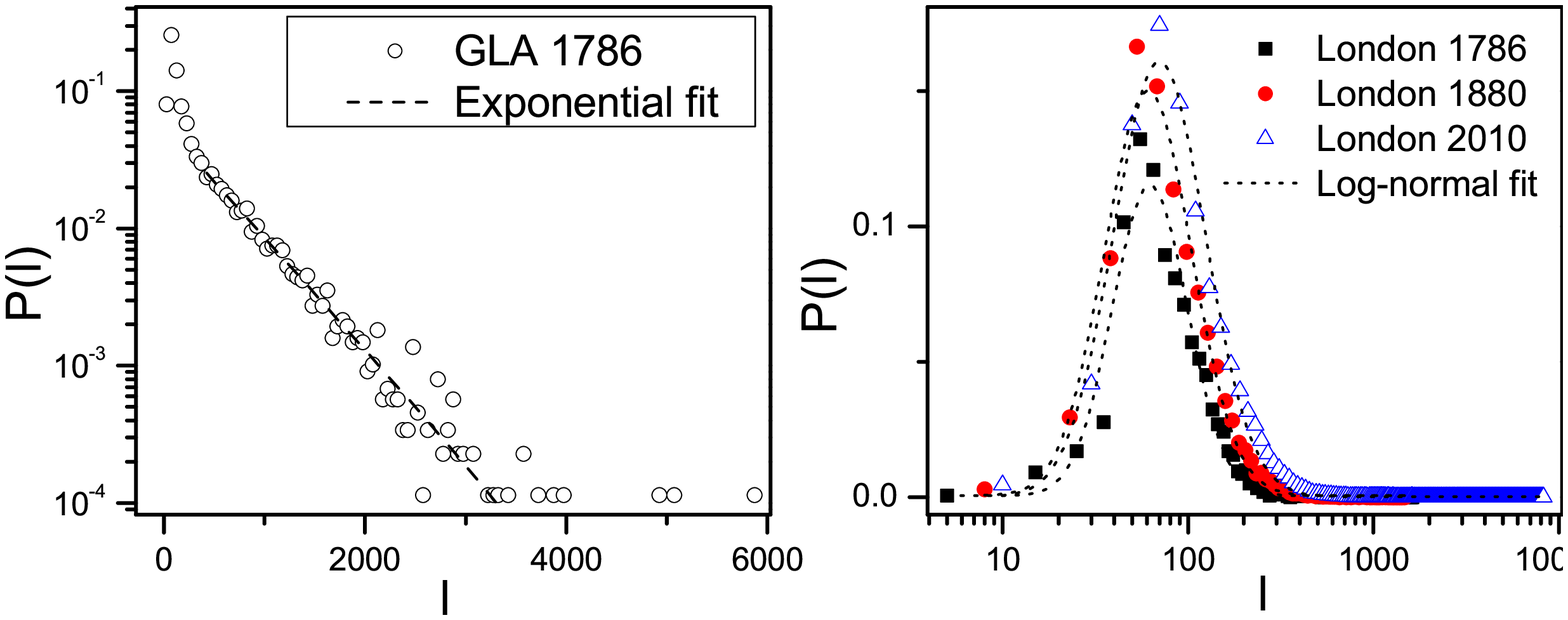}}
\centerline{ \includegraphics[width=0.5\textwidth]{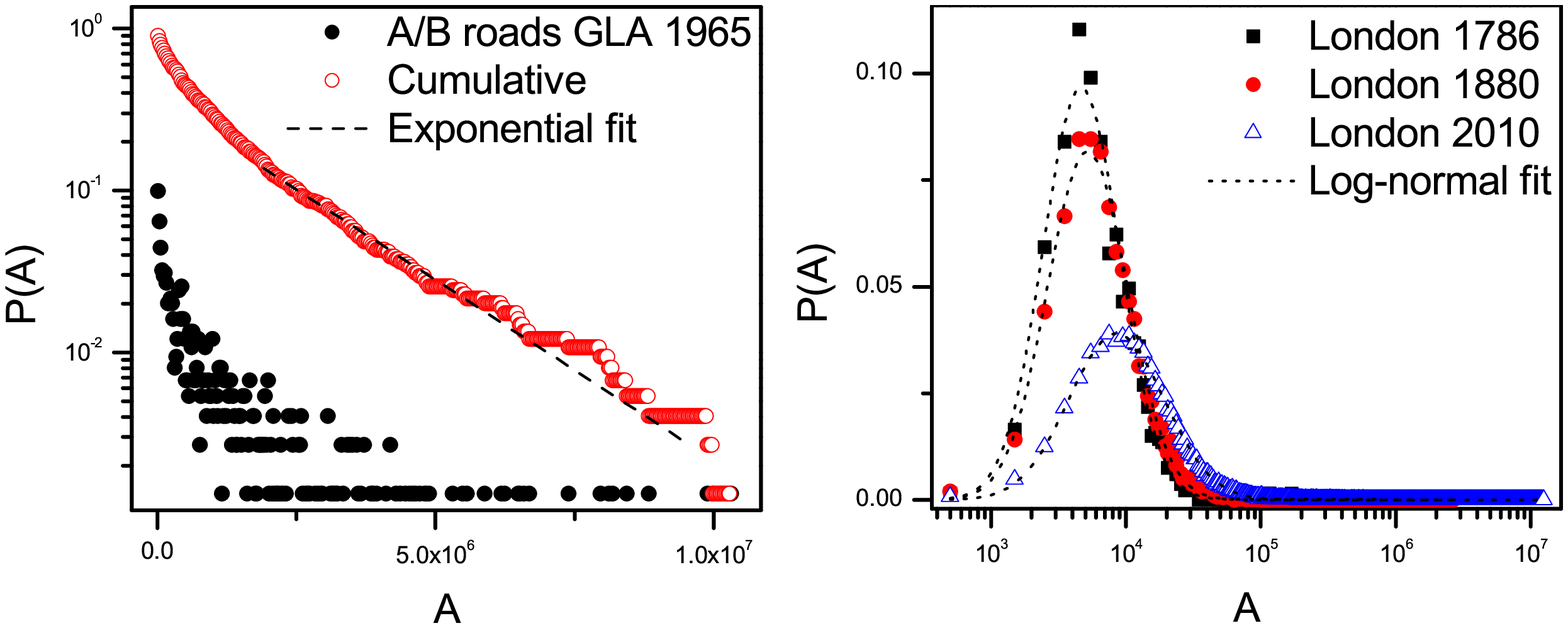}}
  \centerline{\includegraphics[width=0.5\textwidth]{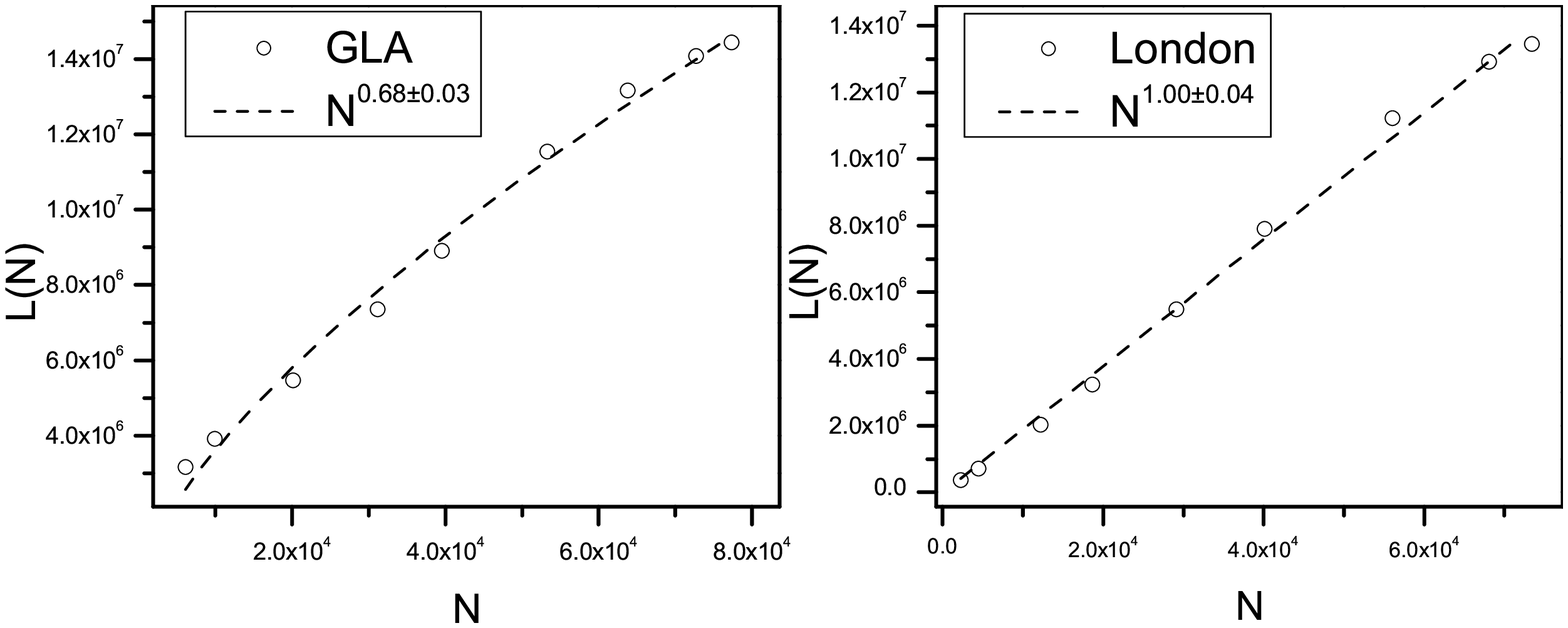}}
 \caption{\label{f7} Top-left panel: Street length distribution as measured in the GLA area in 1786 with exponential fit ($R^2=0.99$). Top-right panel: Street length distribution for   London as defined by the Jenks' algorithm about one century apart with lognormal fit ($R^2=0.97/0.99/0.99$ for 1786, 1880, 2010). Middle-left panel: Parcel area distribution for the network generated by motorways, A and B roads in the 1965 GLA area  with exponential fit ($R^2=0.99$). Middle-right panel:  Parcel area distribution for London with lognormal fit ($R^2=0.98/0.99/0.99$ for 1786, 1880, 2010). Bottom-left panel:  Total street length as a function of the number of intersections for the GLA area with allometric fit ($R^2=0.99$). Bottom-right panel:  Total street length as a function of the number of intersections for London with allometric fit ($R^2=0.99$). }
\end{center}
 \end{figure}

In order to split the core from the wider area, by separating the area of high intersection density patterns from the parts of the maps where the intersection density is low, we apply the \textit{Jenks natural breaks algorithm} to the GLA density maps \cite{jenks}. This methodology begins with a certain number of classes, then generates regions minimizing the within-class variance, while maximizing the inter-class variance between different regions. Setting the number of classes for the intersection density equal to two, we find that the algorithm clearly identifies urbanized areas from areas that are not urbanized. We then use the boundaries of these regions as those of the city. In this way, the spread of urban development within the GLA area is  well-defined and the  boundary for London's city core  is clearly demarcated through the years, as we show in Fig.\ref{f5}. 

We refer to the urban core areas shown in  Fig.\ref{f5} as ``\textit{London}", in order to distinguish it from the GLA area in Fig.1. Such a choice enables us to recognise the emergence of some robust network properties that identify the city as a well-defined physical object. Street length and face area distributions are robust features emerging from our city boundaries definition.  These properties are shown in Fig.\ref{f7}.

In the top panels of Fig.\ref{f7}, we show the street length distribution $P(l)$  for the street segments, defined by every two consecutive intersections. In the top-left panel, the measure is calculated for the GLA area in 1786, when it is mostly not yet urbanized. The distribution displays a clear exponential tail. In the top-right panel, we show the same measure computed for three time slices which differ by about one century for London. In this case, the distribution is robustly log-normal throughout  the entire 224 years (see $R^2$ values in the caption).

 The city face or parcel area  (referred in common language as city blocks) is a quantity that has been considered in statistical physics, since it displays a fat tail distribution \cite{rap,lammer}. 
  In the middle-right panel of Fig.\ref{f7}, we show the face area distribution $P(A)$ in the 1786, 1880 and 1900 London street networks. The plots are  well fitted by log-normal distributions. 
 This finding might appear to be contradictory to previous claims that such a distribution is scale free \cite{barth2011, masuccilond,  lammer}, but in reality the confusion between lognormal and power law distributions is well known and can be easily understood \cite{lognorm}.
 However, it is quite reasonable to think that the city face area distribution is not a scale-free phenomena, since there is an evident limit for the size of urban blocks, which are rarely smaller than $100m^2$.
  The face area distribution is a robust property of the system as defined by the minor roads.
   To see this, in the middle-left panel of Fig.\ref{f7}, we  show the face area distribution for the 1965 map, when minor roads are excluded. From the cumulative distribution, it is possible to see that the tail of this distribution is exponential. 

In the bottom panels of Fig.\ref{f7}, we measure the total length of the street network $L(N)$ as a function of the number of  intersections $N$.
 In \cite{bartprl,strano}, this measure is shown to be sub-linear, with an exponent close to 0.5, i.e. $L(N)\propto \sqrt{N}$.  
 In the bottom-left panel, we show that measuring this quantity in the GLA area, we also obtain a sub-linear trend, with exponent 0.68. 
 Nevertheless, when we measure this quantity for the London area as defined above, the behaviour is linear (bottom-right panel), showing that urban statistical behaviour is very sensitive to the city boundary definition.
 However this measure is merely qualitative and it  serves us to illustrate how the definition of the city boundary  can drastically change the result of a statistical measure, so that the linear trend has to be considered purely qualitative. 
 In the next section we provide  a better approximation for this quantity.

\section*{Capacitated growth in London}

The next step in our analysis is to examine  the properties of the planar graphs representing London from 1786 to 2010.  In a city, there are structures such as bridges and tunnels which violate the planarity of the graph, but as the percentage of such violations is negligible ($<0.01\%$), the planar graph representation can be considered an excellent approximation.

As we stated in the previous section, the growth of a city can be seen as a percolation phenomena in a two-dimensional space. 
However, in the UK and particularly in London,   urban sprawl is highly controlled by bands of open space  forming the \textit{green belt}, which in London's case was formally incorporated in 1953 \cite{greenbelt}, but dates back as an idea to the time of Queen Elizabeth I.
 We can therefore argue that the elements forming the street network of London, i.e. the intersections and the street segments, grew in time as a space-filling phenomena to the capacitated limit determined by this green belt.
In analytical terms we can say that if $f(t)$ is the number of intersections or street segments defining the network, the simplest growth dynamics for such elements to the  capacitated limit can be expressed as:

\begin{equation}\label{log0}
\frac{df(t)}{dt}=rf(t)\left(1-\frac{f(t)}{C}\right),
\end{equation}

where $r$  is the \textit{growth rate} and $C$  the \textit{carrying capacity}. The carrying capacity represents the bias to the free growth of the system, i.e. when $f(t)=C$, the growth of the network stops ($df=0$). In the case of $C=+\infty$, the solution of Eq.\ref{log0} is an exponential function with growth rate $r$, that is the solution for unbiased growth.
The general solution of Eq.\ref{log0} has been well known since 1838 in the work of Verhulst \cite{logistic}, and it is the \textit{logistic function}:

\begin{equation}\label{log}
f(t)=\frac{C}{1+e^{-r(t-t_0)}},
\end{equation}

where $t_0$  is the inflection point, i.e. $\partial^2 f /\partial t^2|_{t=t_0}=0$. 

\begin{figure}[ht]
\begin{center}
\centerline{\includegraphics[width=0.5\textwidth]{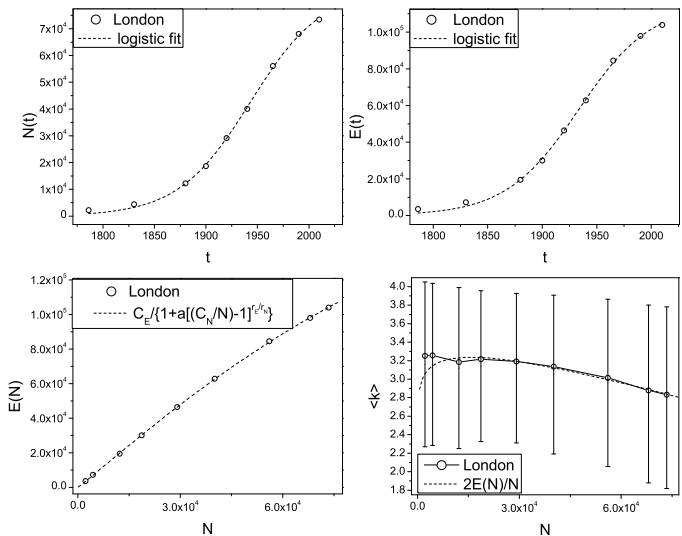}}
 \caption{\label{f9} Top-left panel: The number of street intersections $N(t)$ as a function of time and the logistic fit ($R^2=0.999$). Top-right panel: The number of street segments $E(t)$ as a function of time and the logistic fit ($R^2=0.998$). Bottom-left panel: The number of street segments $E(N)$ as a function of the number of street intersections ($R^2=0.999$). Bottom-right panel: The average degree $<k(N)>$ of the networks as a function of the number of street intersections ($R^2=0.74$).}
\end{center} 
 \end{figure}

In the top panels of Fig.\ref{f9}, we present the measures of the number of vertices  $N(t)$ and the number of edges $E(t)$  as a function of time $t$. The  goodness of fit of such plots with the logistic function of Eq.\ref{log} is impressive ($Adj.R^2=0.9988$ for $N$, $Adj.R^2=0.9984$ for $E$ ).

  First, Eq.\ref{log} allows us to forecast the asymptotic value for the number of intersections as $N^{\infty}=C_N\approx 85123$, for the street segments  $E^{\infty}=C_E\approx 115615$ and for the related average street connectivity  $<k>^{\infty}=2E^{\infty}/N^{\infty}\approx 2.72$ in London.
  
In the bottom-left panel of Fig.\ref{f9}, we show the growth of the number of links or segments as a function of the number of vertices, i.e. the topological growth of the planar graph. This can be analytically expressed by combining the two logistic functions expressing the growth of the vertices and the edges, and can be written as:

\begin{equation}\label{EN}
E(N)=\frac{C_E}{\left[1+a\left(\frac{C_N}{N}-1\right)^\frac{r_E}{r_N}\right]},
\end{equation}
 where     $a=\exp[r_E(t_{0E}-t_{0N})]$ is constant and $r_E/r_N\approx 1.07$ is close to unity. 

In the bottom-right panel of Fig.\ref{f9}, we show the average degree $<k(N)>$  of the network as a function of the number of vertices $N$. This is a slightly decreasing function of time, indicating that the city tends to become a more \textit{tree-like} structure. Considering that $<k(N)>=2E/N$, this is readily computed from Eq.\ref{EN}.
This is an important result, since the average degree is related to the \textit{treeness} or \textit{loopiness} of the graph \cite{magnascopo}. In particular $<k>=2$ characterizes tree-like structures, while $<k>=4$ characterizes grid-like structures \cite{bollo}. In general in nature we find situations where the average degree is  between these two values. It has been shown that for leaf venation networks, the loopy structure evolves from a tree \cite{melville} and relevant research has been done on the robustness of loopy architectures \cite{banavar, bohn, bohn2007}.
 Here we show that the London's urban street network tends to evolve from a more loopy architecture to a more tree-like one. 
 This is in accordance with a space-filling principle where first a loopy architecture emerges in order to assure a proper and robust circulation of goods in the territory and then empty spaces are filled with leaves.
  This observation is related to the tendency of planned cities to evolve into self-organized cities,  a process represented by the transformation of many European urban cores from rigid Roman grids into more organic structures during the medieval period \cite{Kost}.

\begin{figure}[ht]
\begin{center}
\centerline{\includegraphics[width=0.5\textwidth]{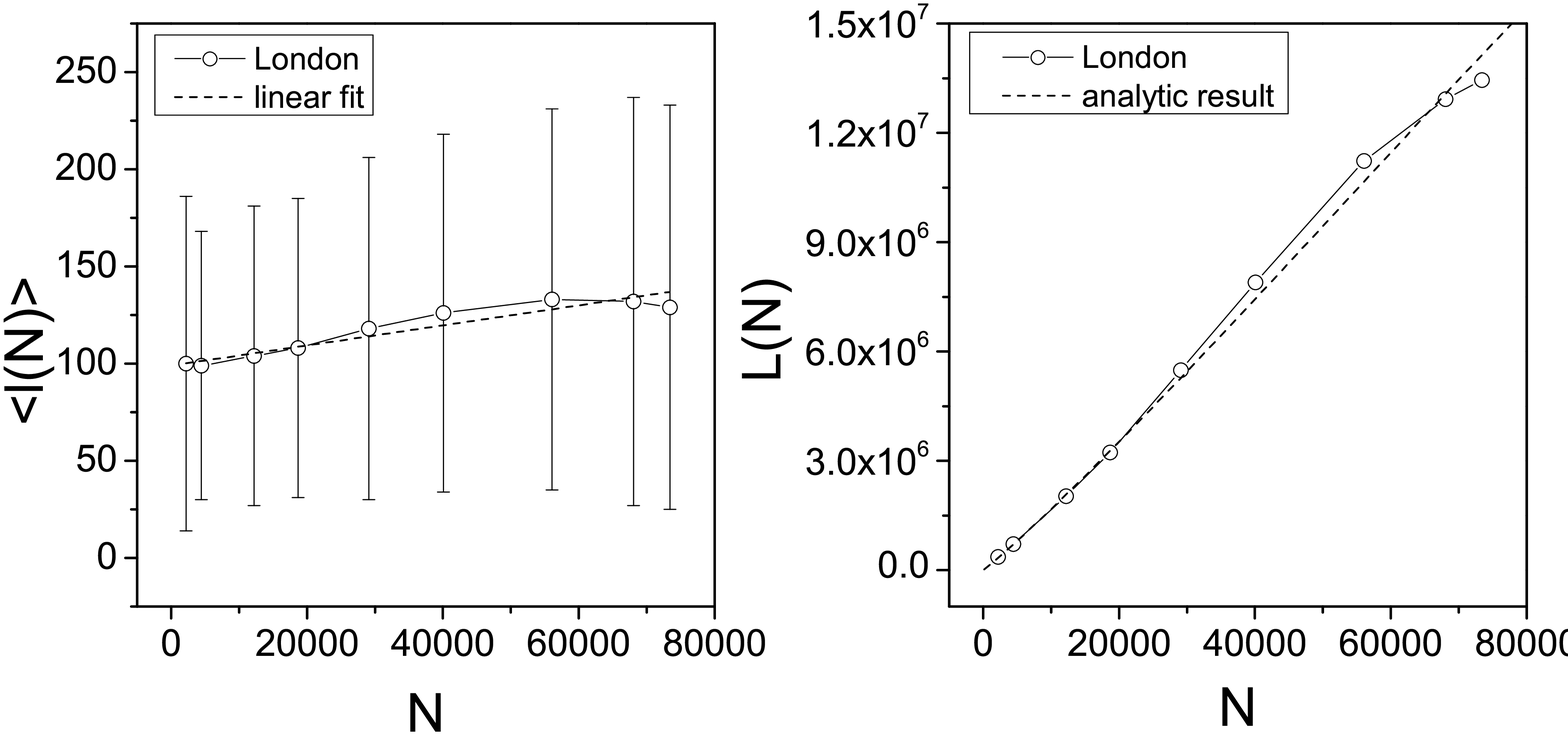}}
 \caption{\label{ltot} Left panel: the average length $<l(N)>$ of the street segments in London's urban core  as a function of the number of vertices. The fitting function is a line, $<l(N)>= 5\cdot 10^{-4}N+99$, (adj. $R^2=0.90$). Right panel: the total length of the network $L(N)$ as a function of the number of vertices, fitted by Eq.\ref{app} ($R^2=0.995$).}
\end{center} 
 \end{figure} 
 
Now we have the instruments to approximate a solution for the total length of the London 	street network $L(N)$ as a function of the number of nodes $N$. Since the length distribution of the street segment is lognormal (Fig.\ref{f7}), the average length $<l>$ of the street segments is a well defined quantity. Therefore the total length of the network can be expressed as 
\begin{equation}\label{app}
L(N)=E(N)<l(N)>,
\end{equation}
 where $<l(N)>$ is a slowly varying function of the number of vertices and can be approximated by a line (see left panel of Fig.\ref{ltot}) and $E(N)$ is given by Eq.\ref{EN}. In the right panel of Fig.\ref{ltot}, we show the total length of the network fitted by Eq.\ref{app}. Again the goodness of the fit is highly satisfactory ($R^2=0.9951$).

  \begin{figure}[ht]
\begin{center}
\centerline{\includegraphics[width=0.5\textwidth]{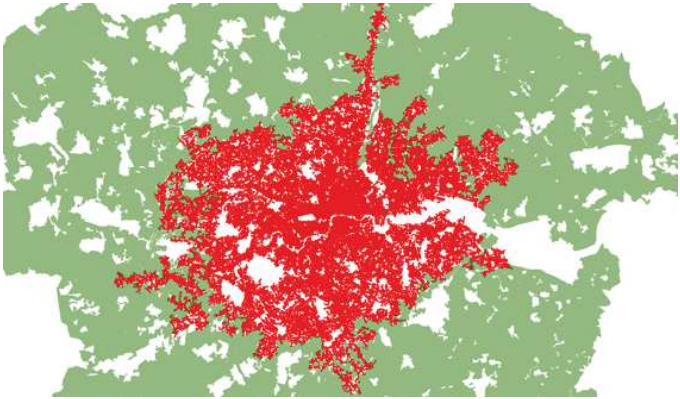}}
 \caption{  \label{f10} The greenbelt surrounding London and the actual extension of London as calculated with the method introduced in the paper. The extensions of London's street network which appear to violate the green belt boundary are in fact filling in highly fragmented interstitial areas not included in the green belt.}
\end{center} 
 \end{figure} 
 
   \begin{figure}[ht]
\begin{center}
\centerline{\includegraphics[width=0.5\textwidth]{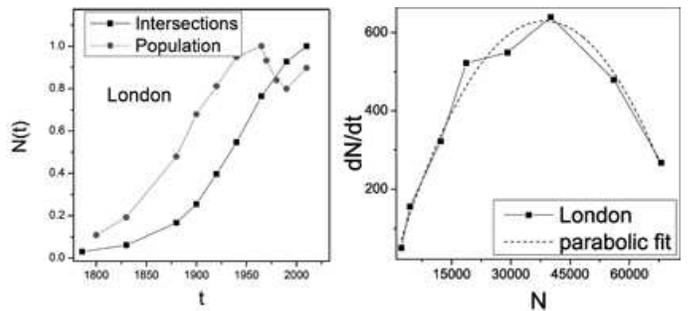}}
 \caption{  \label{f11} In the left panel, we show the street intersection growth $N(t)$ and the population growth for London as a function of time. To appreciate the correlations between the two phenomena we normalised both measures to unity. In the right panel, we show the variation of street intersections during time $dN/dt$ as a function of the number of intersections $N$. This represents a strong violation of  Gibrat's law.}
\end{center} 
 \end{figure}

\section*{Discussion}

Urban street networks are complex phenomena, in the sense that they are  assemblages of elemental units such as streets, which emerge over centuries or millennia, and despite the absence of a unique urban plan, they show global universal statistical properties. 
 Here we show that despite the complexity of urban systems, there are some useful perspectives on the city that consider urban growth to be an analytically tractable phenomenon. 
 Using London as a case study, we have shown that the topological and metrical properties characterizing the complexity of a city can be disentangled and described by a small number of parameters that generate the growth of infrastructure within a limited space. 
 This allows us to derive analytically a few key quantities and to forecast the future evolution of the street network. Also we are able to derive global properties of the network, such as its loopiness, by tracking the time evolution of its elemental constitutive elements.
It is important to underline the fact that the growth of London is strongly influenced by the Green Belt policy, which was first suggested nearly 100 years ago, with policies effectively constraining the city's urbanization process over this period.
 We can also speculate that the introduction of the formal policy in 1953 corresponds roughly to the inflection points on the logistic functions shown in Fig.\ref{f9}, implying that city shapes and forms are highly influenced by local policies.

Regarding the generality of the presented results, we underline that at this stage of the research our generalization of the results is entirely speculative. 
We show how London's growth is highly constrained by local policies, such as the Green Belt.
One could argue that this makes London a special case, but this is not the case at least for England, where all large cities are surrounded by such constraining green areas \cite{gb}. 
As we pointed out earlier, other than England many other cities and city states in the world are similarly constrained.
Moreover we have to consider that a green belt acts as a natural barrier, such as a sea, a lake, a mountain, etc. and that many cities lacking  administrative barriers have natural barriers to constrain their growth.
This is just to say that even if each city is unique,   London's capacitated growth must not be considered as extreme or idiosyncratic.

Another possible critique to this analysis could be the validity of the application of the Verhulst model of Eq.\ref{log0} to the London's street intersection and  segment growth.
As a matter of fact, sigmoidal functions are widely recognisable in nature and can be subject to different interpretations.
Nevertheless, Eq.\ref{log0} represents a population growth to a capacitated limit, that seems to represent our case study and we can see from the left panel of Fig.\ref{f11} that the street intersection growth is strictly correlated with the population growth, where the Verhulst model is first introduced for population growth to a capacitated limit.
Moreover in the right panel of Fig.\ref{f11}, we show that Eq.\ref{log0} can be verified empirically, where $dN/dt$ has a neat quadratic behaviour as a function of $N$. 
Interestingly, this represents a clear and strong violation of Gibrat's law, although we do not formally test this speculation with respect to its measurement. 
 
One could argue that the condensation phenomena that is evident in Fig.\ref{f9} is a result of  limiting the dataset to the GLA area and that if one would consider a wider area a different behaviour could possibly be found. To test this we show in Fig.\ref{f10} the actual greenbelt and the extension of the modern London extracted with the methodology outlined in this paper. The figure demonstrates that the 2010 extent of London as  defined in our methodology does not violate the greenbelt. This supports the argument that the greenbelt effectively stopped the expansion of London's road infrastructure.

Understanding urban growth, particularly the capacitated growth of large cities like London, is central to many perspectives on how we must design and manage urban areas in order to accommodate a sustainable environment.
The analysis of spatially constrained network growth becomes particularly relevant in the light of the ongoing global debate about sustainable development and public policies aimed at limiting suburban sprawl. By the middle of the twenty-first century, the world's urban population is expected to double, increasing from approximately 3.4 billion in 2009 to 6.4 billion in 2050 \cite{UN2011}. This growth will put an enormous strain on natural resources and one of the most popular mitigation measures adopted by planning authorities throughout the globe is to set physical limits to urban expansion \cite{Hall}. Many fast growing world cities (e.g., S\~ao Paolo, Seoul, Beijing, Hong Kong, etc.) have followed London's example and instituted green belts to stop urban sprawl. In the last few decades, a number of metropolitan areas in the US (e.g. Portland, Seattle, etc.) have introduced urban growth boundaries and this policy is being adopted by many cities around the globe. In this light, it is critical to examine the behaviour of infrastructure and transport networks under the conditions of constrained growth as key determinants of urban growth patterns. Our analysis is aimed at shedding light on the spatial behaviour of such systems using London as an important and illuminating case study.

 We consider the observations presented here to be a first step towards a fuller understanding of pattern formation in the evolution of such cities, and it is thus essential that studies of different cities are now needed to explore the existence of more universal properties of such limits on urban growth.
We also speculate that the analytical techniques presented in this paper could be successfully applied in the analysis of other reticulated planar networks, such as leaf venation patterns, circulatory or river networks.

\section*{Acknowledgments}
APM was partially funded by the EPSRC
SCALE project (EP/G057737/1) and MB by the ERC
MECHANICITY Project (249393 ERC-2009-AdG).
We would like to thank the anonymous reviewer that helped us to improve the quality of this paper. 
\bibliography{lond}

\end{document}